\documentclass[apj]{emulateapj}

\usepackage{graphicx,threeparttable}
\usepackage{amsmath}
\usepackage{amssymb}
\usepackage{amsthm}
\usepackage{graphicx}
\usepackage{natbib,color}
\usepackage{multirow}
\usepackage{epsfig}

\newcommand{\teff}{\mbox{$T_{\rm eff}$}}
\newcommand{\teq}{\mbox{$T_{\rm eq}$}}
\newcommand{\logg}{\mbox{$\log g$}}
\newcommand{\vsini}{\mbox{$v \sin i$}}
\newcommand{\mictrb}{\mbox{$v_{\rm t}$}}
\newcommand{\mactrb}{\mbox{$v_{\rm mac}$}}

\newcommand{\halpha}{\mbox{$H_{\rm \alpha}$}}

\newcommand{\msun}{M$_\odot$}
\newcommand{\kms}{\mbox{km\,s$^{-1}$}}
\newcommand{\rhostar}{$\rho_\star$}
\newcommand{\rhosun}{\mbox{$\rho_\odot$}}
\newcommand{\rhoj}{\mbox{$\rho_{\rm J}$}}
\newcommand{\rhopl}{\mbox{$\rho_{\rm pl}$}}
\newcommand{\rj}{R\ensuremath{_{\rm Jup}}}
\newcommand{\mj}{M\ensuremath{_{\rm Jup}}}
\newcommand{\mearth}{M\ensuremath{_{\earth}}}
\newcommand{\rsun}{R$_\odot$}
\newcommand{\rpl}{\mbox{$R_{\rm pl}$}}
\newcommand{\mpl}{\mbox{$M_{\rm pl}$}}
\newcommand{\rstar}{\mbox{$R_\star$}}
\newcommand{\mstar}{\mbox{$M_\star$}}

\def\secos{$\sqrt{e} \cos \omega$}
\def\sesin{$\sqrt{e} \sin \omega$}
\def\feh{[Fe/H]}

\shorttitle{HoSTS: WASP-13}
\shortauthors{G\'omez Maqueo Chew et al.}

\begin{document}

\title{The Homogeneous Study of Transiting Systems (HoSTS) I. The Pilot Study of WASP-13}

\author{
Yilen G\'omez Maqueo Chew\altaffilmark{1,2,3*,4},
Francesca Faedi\altaffilmark{2,3*},
Phillip Cargile\altaffilmark{1}, 
Amanda P.\ Doyle\altaffilmark{5},
Luan Ghezzi\altaffilmark{6,7},
S\'ergio Sousa\altaffilmark{8},
Susana C.\ C.\ Barros\altaffilmark{9},
Leslie Hebb\altaffilmark{1},
Katia Cunha\altaffilmark{6,10},
Simon C.\ Schuler\altaffilmark{11},
Verne V.\ Smith\altaffilmark{11,6},
Andrew Collier Cameron\altaffilmark{12},
Don Pollacco\altaffilmark{2,3*},
Nuno C.\ Santos\altaffilmark{8},
Barry Smalley\altaffilmark{5},
Keivan G.\ Stassun\altaffilmark{1,13} 
}

\altaffiltext{1}{Department of Physics \& Astronomy, Vanderbilt University, Nashville, TN 37235, USA;
yilen.gomez@vanderbilt.edu}
\altaffiltext{2}{Department of Physics, University of Warwick, Coventry CV4 7AL, UK}
\altaffiltext{3*}{Astrophysics Research Centre, Queen's University Belfast, University Rd.\ Belfast, BT7 1NN, UK; part of the work was completed while at QUB.}
\altaffiltext{4}{Centro de Radioastronom\'ia y Astrof\'isica, UNAM, Apartado Postal 3-72, 58089 Morelia, Michoac\'an, M\'exico}
\altaffiltext{5}{Astrophysics Group, Keele University, Staffordshire ST5 5BG, UK}
\altaffiltext{6}{Observat\'orio Nacional, Rua Gal. Jos\'e Cristino 77, Rio de Janeiro, RJ 20921-400, Brazil} 
\altaffiltext{7}{Laborat\'orio Interinstitucional de e-Astronomia - LIneA, Rua Gal. Jos\'e Cristino 77, Rio de Janeiro, RJ 20921- 400, Brazil}
\altaffiltext{8}{Centro de Astrof\'isica, Universidade do Porto, Rua das Estrelas, 4150-762 Porto, Portugal}
\altaffiltext{9}{Aix Marseille Universit\'e, CNRS, LAM (Laboratoire d'Astrophysique de Marseille) UMR 7326, 13388, Marseille, France}
\altaffiltext{10}{Steward Observatory, University of Arizona, Tucson, AZ 85721, USA}
\altaffiltext{11}{National Optical Astronomy Observatory, 950 North Cherry Avenue, Tucson, AZ 85719 USA}
\altaffiltext{12}{School of Physics and Astronomy, University of St Andrews, St Andrews, Fife KY16 9SS, UK}
\altaffiltext{13}{Department of Physics, Fisk University, Nashville, TN 37208, USA}

\begin{abstract}
We present the fundamental stellar and planetary properties of the transiting planetary system 
WASP-13 within the framework of the Homogeneous Study of Transiting Systems (HoSTS). 
HoSTS aims to derive the fundamental stellar (\teff, \feh, \mstar, \rstar), 
and planetary (\mpl, \rpl, \teq) physical properties of known transiting planets using a 
consistent methodology and homogeneous high-quality dataset. 
Four spectral analysis techniques are independently applied to a 
Keck+HIRES spectrum of WASP-13 
considering two distinct cases: unconstrained parameters, and constrained \logg\ from transit light
curves. 
We check the derived stellar temperature against that from a different temperature 
diagnostic based on an INT+IDS \halpha\ spectrum. 
The four unconstrained analyses 
render results that are in good agreement, and provide an improvement of 50\% 
in the precision of \teff, and of 85\% in \feh\ with respect to the WASP-13 discovery paper.
The planetary parameters are then derived via the 
Monte-Carlo-Markov-Chain modeling of the radial velocity and light curves, 
in iteration with stellar evolutionary models to derive realistic uncertainties.
WASP-13 (1.187$\pm$0.065\msun; 1.574$\pm$0.048\rsun) hosts a Saturn-mass, transiting planet (0.500$\pm$0.037\mj; 1.407$\pm$0.052\rj), 
and is at the end of its main-sequence lifetime (4-5.5~Gyr).  
Our analysis of WASP-13 showcases that both a detailed stellar characterization, 
and transit modeling  
are necessary to well determine the fundamental properties of planetary systems, which are paramount
in identifying and determining empirical relationships between transiting planets and their hosts.
\end{abstract}



\section{Introduction}

The detection and characterization of a large number of extrasolar planets with a variety of physical
properties, in different environments and with a range of ages is necessary for the understanding
of the formation and evolution of planetary systems. It is only with precise measurements of the
fundamental properties of the exoplanets and their host stars that the planetary bulk composition can 
be inferred and the planetary structure 
probed, and thus, we can explore the underlying physical processes involved in their
formation and evolution.  

Transit surveys, such as SuperWASP \citep{Pollacco2006}, have been extremely successful
in discovering planets for which measurements
of their masses and radii are possible.  
These have revealed a large diversity of physical properties of the extrasolar planets
and their host stars (see e.g., \citealt{Baraffe2010}).
 With more than 290 transiting exoplanets confirmed to date\footnote{See http://exoplanet.eu}, it is now possible to conduct statistical studies of  planetary properties, and thus, derive more robust  
empirical relationships between the planets and their host stars.  
For example, it is generally thought that in the case of Hot Jupiters  
 the planetary
radius is correlated with the planet equilibrium temperature and the stellar irradiation, and
anticorrelated with stellar metallicity 
\citep[e.g.,][]{Santos2004,Guillot2006,Laughlin2011,Enoch2012,Faedi2011,Demory2011}. 
\citet{Buchhave2012}, using recent Kepler results, find that giant planets are found around metal-rich stars, 
while those with smaller radii than four times that of the Earth are found to orbit stars with a large range in metallicity 
(-0.6 $\le$ \feh\ $\le$ 0.5 dex).
This is compatible with previous observational results \citep{Udry2007,Sousa2008,Sousa2011b,Ghezzi2010a}, 
which show that Neptunian planets do not form preferentially around metal-rich stars. 
Moreover,  
\citet{Adibekyan2012a,Adibekyan2012b} show that although the terrestrial planets can be found in a low-iron regime, 
they are mostly enhanced by alpha elements as compared to stars without detected
planets showing that metals continue to be important also for the formation of these planets.

This paper presents the pilot study of our project, 
entitled Homogeneous Study of Transiting Systems (HoSTS), that will derive a homogeneous
set of physical properties for all transiting planets and their hosts stars, aiming to minimize 
the effects of any systematics in the measurements due to quality of the data and/or technique applied. 
Individual studies of single planetary systems 
make systematic uncertainties difficult to identify, and quantify. For example, \citet{Mancini2012} found 
HAT-P-8b to have a radius $\sim$14\% smaller than previous estimates (a difference larger than the quoted uncertainties),
making it consistent with other transiting planet radii, and not significantly inflated. 
As more planets are being discovered and/or re-analyzed, the observed trends, like the anomalously large planetary radii, remain to be confirmed.    
A number of recent studies employing consistent analysis procedures  
for subsets of the known stars with transiting planets have been attempted 
\citep[e.g.,][]{Torres2008,Ammler2009,Southworth2008,Southworth2009,Southworth2010,Southworth2011,Southworth2012};
however, these have largely employed heterogeneous spectroscopic datasets and 
adopt the stellar properties, like \teff\ and \feh, from the literature.
Because these measurements are non-homogeneous
---arising from different spectroscopic analysis techniques applied to spectra obtained with different spectrographs, different resolution, etc.---
the typically quoted uncertainties of $\sim$10\% in the published stellar
and planetary  mass and radius 
likely contain currently uncharacterized systematics. 
More recently, \citet{Torres2012} have thoroughly 
analyzed new and archival spectra (from different instruments, and with varied signal-to-noise ratio and resolution) 
of 56 transiting planet hosts  
comparing three different stellar characterization methods.  
\citet{Torres2012}  focused on the stellar hosts, deriving a new set of homogeneous 
spectroscopic stellar properties  and have been able to identify systematic errors due to the stellar characterization techniques
applied. 
Thus, any empirical trend identified among the 
physical properties, like the observed inflated radii of hot Jupiters with respect to planetary models,
may have to be revised.

HoSTS extends these previous studies in that the stellar host properties are derived from 
a homogeneous, high-quality spectral dataset applying four stellar characterization techniques, 
{\it and} that the planetary properties are also derived consistently. 
By means of our homogeneous spectral dataset and subsequent analyses, we will be able to investigate systematic uncertainties on the derived stellar properties arising not only from the methodology,
as exemplified by \citet{Torres2012}, but also from the quality of the data.  
We will combine iteratively our results with the best available radial velocity data and transit photometry in the literature to derive a homogeneous set of properties for the transiting systems.  The resulting consistent set of physical properties will allow us to further explore known correlations, e.g., core-size of the planet and stellar metallicity, and to newly identify subtle relationships providing insight into our fundamental understanding of planetary formation, structure, and evolution.  And thus, 
this will allow us to reevaluate the planetary properties, of each planet alone, and with respect to different planet populations. 

In this paper, we present our HoSTS pilot study of the transiting system WASP-13 which
is composed of a Saturn-mass planet around an early G-type star. Section \ref{data}  
describes both the data acquired by our team and the data from the literature
utilized in our analyses.  Section~\ref{analysis} 
describes the spectral analysis of the WASP-13 spectra 
implementing four different techniques 
in order to derive the stellar spectroscopic properties. 
For each of the four stellar characterization methods, 
we present two different cases for which stellar properties have been derived: 
a) an unconstrained analysis where all parameters are left free; and b) applying
an external constraint on \logg. 
We have used the temperature diagnostic based on \halpha\ to verify the values derived
from the stellar characterization methods, as well as the effect of fixing \teff\ on the other spectroscopically determined stellar properties.  
Then we describe the modeling of the system's radial velocity and light curves to derive 
the stellar and planetary properties.
In section \ref{concl}, we discuss the results from our pilot study of WASP-13, and outline the future work of the HoSTS project. 

\section{Data}\label{data}

\begin{figure*}
\includegraphics*[width=0.49\textwidth]{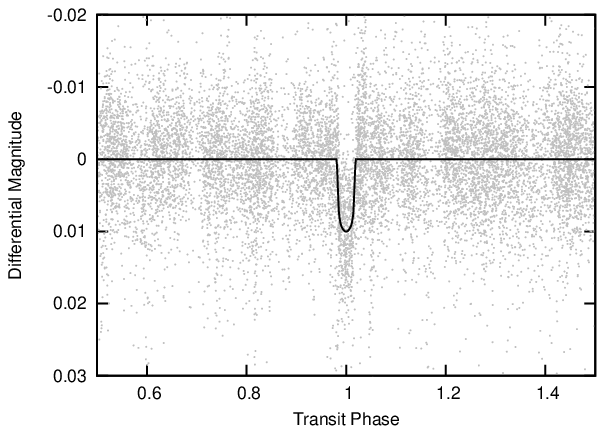}
\includegraphics*[width=0.49\textwidth]{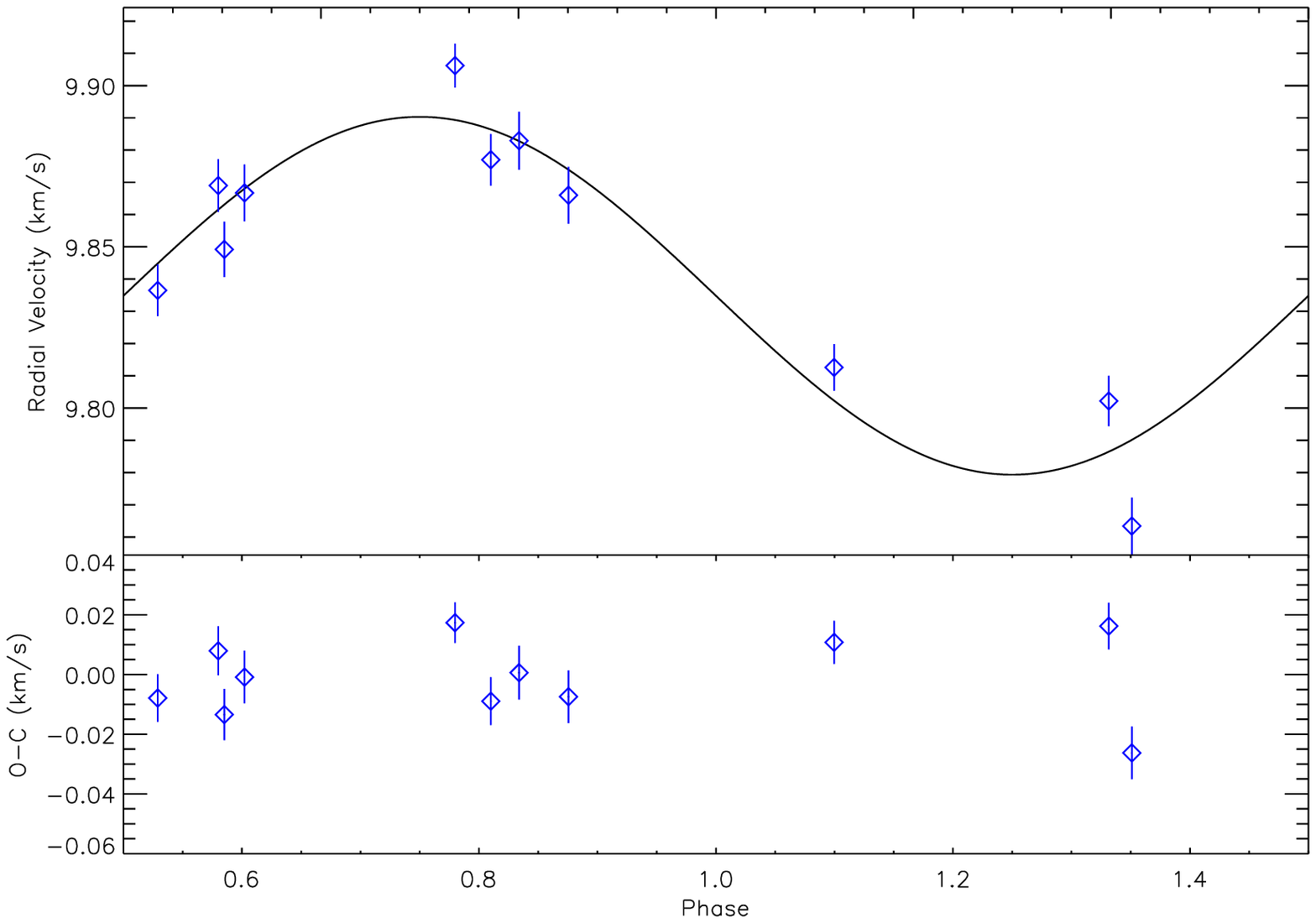}
\caption{\label{swasp}\label{rvs} 
{\it Left}: SuperWASP-N Light Curve of Transiting Planetary System WASP-13. 
Comprising over 12,100 data points and spanning from 2006 November to 2009 April,  
the SuperWASP light curve is shown with gray points with the model transit light curve (see \S\ref{mcmc}) in black.
The photometric data has been folded over the orbital period, and exhibits at phase zero the characteristic dip in 
brightness as the Saturn-mass planet WASP-13b  (\mpl\ = 0.50 $\pm$ 0.01 \mj) 
passes in front of its host star blocking part of the stellar light  
every 4.353~days.
The median photometric uncertainty of the light curve is 0.006 mag.
{\it Right}:
The radial velocity measurements from the SOPHIE instrument
at the Observatoire d'Haute Provence \citep{Skillen2009} 
are shown against the our
model radial velocity curve that describes the reflex motion of WASP-13 due to the 
presence of the planet.  The residuals to the fit are shown in the lower panel.   
}
\end{figure*}

\begin{figure}[!ht]
\includegraphics*[width=.95\columnwidth]{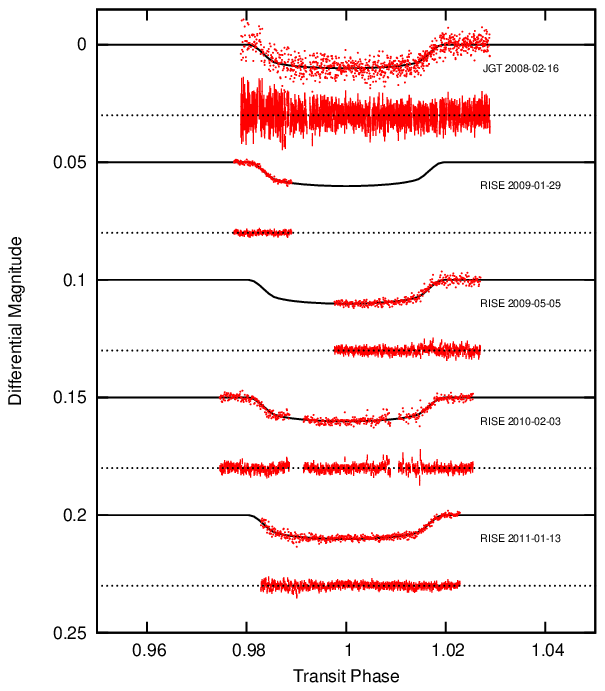}
\caption{\label{lcs} WASP-13b Follow-up Transit Light Curves.  
The transit light curves of the WASP-13 system
from JGT \citep{Skillen2009} and RISE \citep{Barros2012} 
are shown in red points overplotted with the model light curves (in solid black lines) from our MCMC analysis (\S\ref{mcmc}), 
with their corresponding residuals including the uncertainty of the photometric data directly below (red errorbars).  
The light curves and residuals have been shifted vertically from zero for clarity.   
} 
\end{figure}

For our analysis of the
planetary system WASP-13, we have acquired new spectroscopic data described below 
in \S\ref{hires} and \S\ref{int}, and  
have utilized the data available from the literature and the SuperWASP archive.
The SuperWASP light curve includes over 12,100 data points
observed with the SuperWASP-North facility in La Palma, Spain \citep{Pollacco2006} from
2006 November to 2009 April (see Fig.~\ref{swasp}, left panel), 
These data expand the span of the SuperWASP light curve
from that of the discovery paper for two additional years, 
and have a median photometric uncertainty of 0.006 mag.  
From \citet{Skillen2009}, we have obtained the system's radial velocities 
acquired with the SOPHIE instrument mounted on the 1.9-m telescope 
at the Haute Provence Observatory (see Fig.~\ref{rvs}, right panel),
as well as the James Gregory Telescope (JGT) differential 
photometry in the $R$-band.   
Additionally, we have adopted  
the high-cadence, high-precision light curves observed with the RISE instrument
on the Liverpool Telescope published by \citet{Barros2012}.   
All follow-up light curves are shown in Fig.~\ref{lcs}.

\subsection{WASP-13 \halpha\ Long-slit Spectrum}\label{int}
For this work, we obtained a long-slit spectrum of the planet host WASP-13 around the \halpha\ line  
using the Intermediate Dispersion Spectrograph (IDS) 
mounted on the 2.5-m Isaac Newton Telescope at the Roque
de los Muchachos Observatory in La Palma, Spain. 
We used the H1800V grating with the RED+2 CCD, and a 1\farcs4 slit yielding a dispersion of 0.35 \AA/pixel and a  
resolution of R$\sim$10\,000 at 6560\AA. 
The observations of WASP-13 and the standard calibrations,
including a spectrum of Vega and Arcturus, were taken on 2012 May 08 and 09.  
To allow for a precise measurement of the \halpha\ profile, the WASP-13 spectrum 
had a signal-to-noise ratio of $\sim$500  
as calculated by the IDL function {\tt DER\_SNR}.  
The derived temperature (\S\ref{teff}) is used as an independent check on the 
effective temperature of WASP-13 derived from the stellar characterization methods.

\subsection{WASP-13 HIRES Spectrum}\label{hires}
We observed WASP-13 on  2011 March 14 UT 
with the High Resolution Echelle Spectrograph (HIRES) on Keck-I \citep{Vogt1994}. We observed in the spectrograph's
``red" (HIRESr) configuration with an echelle angle
of -0\fdg{018} and a cross-disperser angle of 0\fdg{737}. 
We used the KV418 
order-blocking filter and the 0\farcs{57}$\times$7\farcs{0} slit, and the  
chip was binned by 2 pixels in the spatial direction during readout. 
The resulting resolving power is R$\sim$72,000. 

We obtained three consecutive integrations of WASP-13, each
of 600 s. ThAr arc lamp calibration exposures were obtained
before and after the WASP-13 exposures, and sequences of bias
and dome flat-field exposures were obtained at the end of
the night. The WASP-13 exposures were processed along with
these calibrations using standard IRAF\footnote{IRAF is distributed by the National Optical Astronomy Observatory, which is operated by the Association of Universities for Research in Astronomy (AURA) under cooperative agreement with the National Science Foundation.} 
tasks and the MAKEE
reduction package written for HIRES by T. Barlow. The latter
includes optimal extraction of the orders as well as subtraction
of the adjacent sky background. The three exposures of WASP-13 
were processed separately and then median combined with
cosmic-ray rejection into a single final spectrum. The signal-to-noise
ratio of the final spectrum is $\sim$300 per resolution element.

\section{Analysis}\label{analysis}

In this section, we describe the methods applied to our WASP-13 dataset in order to derive the physical properties
of the planetary system.

\subsection{Determination of \teff\ from \halpha\ Spectrum}\label{teff}
This analysis is based on the \halpha\ spectrum of WASP-13 described in \S\ref{int}. 
The Balmer lines provide an excellent \teff\ diagnostic for stars cooler
than about 8000~K due to their virtually nil gravity dependence
\citep{Gray2008}.  Normalization of the observations is critical,
which the shape of the Balmer line must be preserved
\citep{Smith1988}. The extracted spectrum
was normalized using a low-order polynomial fitted to the continuum regions more
than 100{\AA} either side of \halpha, in order to avoid any distortion due
to the weak wings of this profile. The spectrum was then analysed using {\sc uclsyn} 
with \cite{Castelli1997} ATLAS9 models with no overshooting and
\halpha\ profiles calculated using VCS theory \citep{Vidal1973}.
The best-fitting \halpha\ profile has \teff\ = 5850$\pm$60~K.  

However, the use of Balmer lines as temperature diagnostics is not without its difficulties due to uncertainties caused by different line broadening theories 
\citep{Stehle1999,Barklem2000,Allard2008}, 
the treatment of atmospheric convection \citep{Gardiner1999,Heiter2002}, 
and non-LTE effects \citep{Barklem2007}. 
We, therefore, also fitted the \halpha\ profile in the KPNO solar spectrum  \citep{Kurucz1984} 
which gave a \teff\ 70 $\pm$ 20~K lower than the direct value of 5777~K. Thus, \halpha\
appears to underestimate stellar effective temperatures. 
Adding 70 $\pm$ 20~K to the \teff\ derived above, it gives \teff\ = 5920 $\pm$ 60~K for WASP-13. 
This was recently investigated in detail by \citet{Cayrel2011} who found similar systematic differences,
and provide a correction:  
\[ T_{\rm eff} ({\rm direct}) = 20.3 + 1.014 \times T_{\rm eff} (\halpha)
\]
with an uncertainty of 31~K.
Applying this correction, and adding the errors in quadrature, gives
$T_{\rm eff} = 5950 \pm 70$~K for WASP-13.



Both temperatures from our \halpha\ temperature diagnostic agree with each other, and with 
the \teff\ derived from the infrared flux method \citep[\teff\ = 5935 $\pm$ 183~K;][]{Skillen2009}.
Moreover, they are consistent with the temperatures derived in the stellar characterization methods described below.  

\subsection{Stellar Characterization Analysis \label{starch}}

We apply to the HIRES echelle spectrum  (\S\ref{hires}) four different stellar characterization methods 
that are extensively used in the exoplanet literature.
Each method is done independently from each other and is described in the subsections below (\S\ref{sme}--\ref{porto}).  
Method A is based on the technique of spectral synthesis,  
which compares an observed spectrum to synthetic model spectra generated for a range of stellar parameters. 
The best fitting model (based on a $\chi^2$ minimization) defines
the final atmospheric parameters. 
The other three methods (B, C, and D) are based on the principle of excitation/ionization equilibrium of iron lines,
in which 
equivalent width measurements of many lines are used to determine iron abundance, and the stellar
atmospheric properties.   
Methods B, C, and D are each unique in their choice of linelists, model atmospheres, equivalent width measurements, 
continuum normalization, and convergence criteria. 
Furthermore,
Method B makes an absolute iron abundance measurement of the star, whereas 
the other three methods are differential analyses, and derive a stellar metallicity relative to the Sun.
Method A includes a careful determination of 
the linelist parameters, i.e., excitation potential and oscillator strength, to match the spectrum of the Sun 
using spectral synthesis.
Method C does a differential line-by-line analysis relative to the Sun using the same instrument setup.
While, Method D uses measured equivalent widths and 
a standard Solar iron abundance \citep[e.g., 7.50 $\pm$ 0.04;][]{Asplund2009}   
to derive the line properties.

For each of the four stellar characterization methods, 
we present two distinct cases: 
a) the {\it unconstrained} analysis, where \teff, \logg, and \feh\ (and \vsini\ when appropriate) are derived freely; and 
b) constraining \logg\ from the mean stellar density as
determined from the transit model to derive the other stellar properties. 
The spectroscopically-determined stellar parameters derived for each case above with all four methods are shown in Table~\ref{starchar}.  
Furthermore, the \teff\ derived using a different temperature diagnostic based on an \halpha\ spectrum
is used to check the stellar properties in Table~\ref{starchar}, as well as to  
explore the effect on the derived stellar spectroscopic properties by fixing \teff.  The resulting
spectroscopic properties are shown in Table~\ref{fixteff}. 

Given the quality of our data, and the nuances of each of the stellar characterization methods, the preferred
solution for each of the four methods is that derived through the unconstrained analysis. 
In the last column of Table~\ref{starchar},  we present the weighted mean of the four unconstrained solutions for each method
which are used to derive  
the stellar mass and radius, and the planetary properties (see \S\ref{mcmc}).
Additionally, we report two errors on the  
stellar properties of Table~\ref{starchar}: the first is calculated from the weighted quadrature sum of the individual internal errors, 
and the second is a measure of the systematic uncertainty based on 
standard deviation of the individual measurements.  
The systematic uncertainty is likely to be underestimated in the case of WASP-13, 
because only the four measurements from the unconstrained cases are taken into account. A more realistic 
systematic uncertainty will be possible once a larger HoSTS sample has been analyzed in the 
same consistent manner as we present in this paper.  

\subsubsection{Method A -- SME}\label{sme}
Method A consists on the implementation of Spectroscopy Made Easy \citep[SME v.3.54][]{Valenti1996} to derive stellar
parameters of WASP-13 described below. We base the general method of our SME analysis on that given in
\citet{Valenti2005} including the grid of model atmospheres and derivation of macroturbulence;
however, we use a line list, synthesized wavelength ranges, and abundance pattern adapted from
\citet{Stempels2007,Hebb2009}.  

In general, SME uses the Levenberg-Marquardt (LM) algorithm to solve the nonlinear least-squares  
problem of fitting an observed spectrum with a synthetic spectrum. Like any nonlinear
least-squares algorithm, the LM based solver in SME requires a good initial guess and a smoothly
varying $\chi^{2}$ surface in order to consistently find the absolute global minimum (what we are
calling the optimal solution). In addition, a single SME best-fit solution does not allow for an
estimation of the error in the solution apart from the error calculated from that solution's covariance
matrix. This does not take into account the deviations from the best-fit solution depending on the
specific choice of initial parameter values nor the internal precision of the solver.

We have expanded on the technique outlined in \citet{Valenti2005} that allows us to operate SME in an
automated fashion and explore the effect of the initial conditions on the final resulting stellar parameters.
Using the ACCRE High-Performance Computing Center at Vanderbilt University, we have developed an extensive Monte
Carlo approach to using SME. We start by randomly selecting 500 initial parameter values from a multivariate normal
distribution with 5 parameters: \teff, \logg, \feh, [M/H], and \vsini. For WASP-13, we defined this
distribution using the derived stellar parameters and uncertainties from \citet{Skillen2009}. The microturbulence
(\mictrb) for each of these initial values in this multivariate distribution was fixed at 1.01 $\pm$ 0.17 \kms, estimated using a
polynomial fit (Eq.~\ref{eq:mictrb}) to the HARPS sample of stellar \teff\ and microturbulence \citep{Sousa2011b}
at \teff\ from \citet{Skillen2009}.  The value of \mictrb\ is kept fixed throughout our SME analysis, because the scatter 
of the HARPS sample around this temperature is larger than the change in \mictrb\ in the range of temperatures explored
in all cases.  
Furthermore, the change in the microturbulence value within its uncertainties does not affect significantly  
the derived \feh. 

\begin{equation}\label{eq:mictrb}
\mictrb =  0.909148 + (\teff - 5700)/1318 \\
 + (\teff - 5700)^{2}/1660^2 
\end{equation}

We then allow SME to find a best-fit synthetic spectrum and solve for the free parameters for the full distribution
of initial guesses, producing 500 best-fit solutions for the stellar parameters. We determine our final measured stellar
properties by identifying the output parameters that give the optimal SME solution (i.e., the solution
with the lowest $\chi^2$). The overall SME measurement
uncertainties in the final parameters are calculated by adding in quadrature: 
1) the internal error determined from the 68.3\% confidence region in the $\chi^2$ map, and 
2) the median absolute deviation of the parameters from the 500 output SME solutions to account for the correlation between the initial guess and the final fit. 

Following this procedure, we solved for the parameters of WASP-13 first letting all fitted parameters be free; 
then using a 
constraint on \logg, and letting the other parameters free; and, finally analyzing the spectrum 
fixing \teff. The resulting optimal parameters and uncertainties are given in the first results column of Tables~\ref{starchar} and \ref{fixteff}.

Our choice of the unconstrained analysis as the preferred solution for method A does not follow the
conclusions of  \citet{Torres2012}, where they  
suggest that fixing the \logg\ (case b  above) is the best approach when using synthesis-based methods.  
However, we find that our method A differs in their implementation of SME in the treatment of microturbulence,
the linelist, the sampling of a large parameter space in initial parameters, and the convergence criteria.
Specifically our linelist includes the Na I D region between 5849-5950~\AA, a
gravity and temperature sensitive line, as well as the gravity sensitive Mg b triplet region. 
In similarity to the analyses by \citet{Torres2012} and \citet{Valenti2005}, we did not include the \halpha\ region  
due to the difficulty in normalizing the continuum
for such a broad line in an echelle spectrum.  
A more in-depth comparison
will be possible on the larger HoSTS sample, with which we will be able to state more robustly whether 
we need to constrain \logg\ or not, as well as to explore the dependence of physical properties on the data. 

\subsubsection{Method B -- UCLSYN}\label{keele}
Method B consists in the analysis performed with the spectral synthesis package {\sc uclsyn} (University College London SYNthesis; \citealt{Smith1988}; \citealt{Smith1992}; \citealt*{Smalley2001}) using the methods given in \cite{Doyle2013}.  In general, the surface gravity (\logg) was determined using the ionisation balance of the Fe~{\sc i} and Fe~{\sc ii} lines, as well as from the Ca~{\sc i} line at 6439{\AA} and the Na~{\sc i} D lines. The excitation balance of the Fe~{\sc i} lines was used to determine the effective temperature (\teff). A null dependence was required between Fe abundance and equivalent width in order to ascertain the microturbulence (\mictrb) using the \cite{Magain1984} method. The Fe abundance was determined from equivalent width measurements of several unblended lines, and additional least squares fitting of lines was performed when required. 
The projected stellar rotation velocity (\vsini) was determined by fitting the
profiles of several unblended Fe~{\sc i} lines. A value for
macroturbulence (\mactrb) of 3.0 $\pm$ 0.3 {\kms} was assumed, based on the
calibration by \cite{Bruntt2010}.

The parameters obtained from the analysis are listed in the second column of Tables~\ref{starchar} and \ref{fixteff}. 

\begin{table*}
\begin{center}
\begin{threeparttable}[t]
\caption{Spectroscopically-determined Stellar Parameters of WASP-13} \label{starchar}
\begin{tabular}{lccccc} \hline \hline \\ 
\phn  & A & B  & C & D & Weighted Mean\tnote{*} \\ \hline \\
 \phn  & \multicolumn{4}{c}{Unconstrained}\\ \cline{2-5} \\
\teff\ (K)   	&  6003 $\pm$ 65 	& 5955 $\pm$  75 	& 5919 $\pm$ 30		& 6025 $\pm$ 21 &	5989 $\pm$ 16 $\pm$ 48\\ 
\logg    	&   4.16 $\pm$ 0.08	& 4.13 $\pm$ 0.11 	& 4.02 $\pm$ 0.06	& 4.19 $\pm$ 0.03 &	4.16 $\pm$ 0.03 $\pm$ 0.07 \\ 
$\log A$(Fe) 	&   7.54 $\pm$ 0.06\tnote{$\dagger$}	& 7.60 $\pm$ 0.09 	& 7.54 $\pm$ 0.05\tnote{$\dagger$}	& 7.58 $\pm$ 0.05\tnote{$\dagger$} & 7.56 $\pm$ 0.03 $\pm$ 0.03 \\	
\feh
		& 0.04 $\pm$ 0.05	& 0.10 $\pm$ 0.09\tnote{$\dagger$} 	& 0.04 $\pm$ 0.02 	& 0.08 $\pm$ 0.02 & 0.06 $\pm$ 0.01 $\pm$ 0.03 \\ 
\vsini\ (\kms) 	& 5.79 $\pm$ 0.08	& 5.26 $\pm$ 0.25 		& \nodata	& \nodata &  5.74 $\pm$ 0.08 $\pm$ 0.38	\\ 
\mictrb\ (\kms) &   1.01 $\pm$ 0.17\tnote{$\ddagger$}	& 0.95 $\pm$ 0.10 	& 1.53 $\pm$ 0.09	& 1.28 $\pm$ 0.10 &  	1.27 $\pm$ 0.06 $\pm$ 0.29 \\ 
\hline \\
\phn  &	\multicolumn{4}{c}{Fixing \logg\ = 4.10 $\pm$ 0.04 dex}\\ \cline{2-5} \\
\teff (K)	& 5994 $\pm$ 150	& 5955 $\pm$ 70 	& 5912 $\pm$ 30		& 6048 $\pm$ 63 \\
$\log A$(Fe) 	& 7.55 $\pm$ 0.12\tnote{$\dagger$}	& 7.59 $\pm$ 0.09 	& 7.54 $\pm$ 0.05\tnote{$\dagger$}	& 7.56 $\pm$ 0.07\tnote{$\dagger$}\\ 
\feh		& 0.05 $\pm$ 0.11	& 0.09 $\pm$ 0.09\tnote{$\dagger$} 	& 0.04 $\pm$ 0.02	& 0.06 $\pm$ 0.06 \\ 
\vsini\ (\kms) 	& 5.86 $\pm$ 0.22	& 5.26 $\pm$ 0.25  	& \nodata	& \nodata\\	
\mictrb\ (\kms)	& 1.01 $\pm$ 0.17\tnote{$\ddagger$} 		& 1.00 $\pm$ 0.10 	& 1.53 $\pm$ 0.09	& 1.33 $\pm$ 0.10\\
\hline 
\end{tabular}
\begin{tablenotes}
\item[$*$]{\footnotesize The first error ($\sigma_w$) is derived from the uncertainties in 
the individual measurements ($\sigma_i$),  $\sigma_w = [~\sum{1/\sigma_i}~]^{-1/2}$,
and the second error is 
calculated from the standard deviation of the individual measurements.
} 
\item[$\dagger$]{\footnotesize Derived using the most current value for the solar abundance  of iron, $A \log$(Fe)$_{\sun}$ = 7.50 $\pm$ 0.04 \citep{Asplund2009}.}
\item[$\ddagger$]{\footnotesize The \mictrb\ has been adopted from the empirical relationship described in \S\ref{sme}, and is not included in the weighted mean.}  
\end{tablenotes}
\end{threeparttable}
\end{center}
\end{table*}

\begin{table*}
\begin{center}
\begin{threeparttable}[t]
\caption{Stellar Properties of WASP-13: Fixing \teff\ = 5950 $\pm$ 70 K} \label{fixteff}
\begin{tabular}{lcccc} \hline \hline \\
\phn  & A & B  & C & D  \\ \hline \\
\logg          & 4.14 $\pm$ 0.19       & 4.13 $\pm$ 0.11       & 4.07 $\pm$ 0.13       & 4.06 $\pm$ 0.10 \\
$\log A$(Fe)   & 7.56 $\pm$ 0.07\tnote{$\dagger$}      & 7.60 $\pm$ 0.09       & 7.56 $\pm$ 0.06\tnote{$\dagger$}      & 7.52 $\pm$ 0.07\tnote{$\dagger$} \\ 
\feh           & 0.06 $\pm$ 0.06       & 0.10 $\pm$ 0.09\tnote{$\dagger$}      & 0.06 $\pm$ 0.04       & 0.02 $\pm$ 0.06 \\ 
\vsini\ (\kms) & 5.88 $\pm$ 0.05       & 5.26 $\pm$ 0.25               & \nodata       & \nodata\\     
\mictrb\ (\kms)        & 1.01 $\pm$ 0.17     & 0.95 $\pm$ 0.10       & 1.53 $\pm$ 0.08       & 1.30 $\pm$ 0.10 \\
\hline
\end{tabular}
\begin{tablenotes}
\item[$\dagger$]{\footnotesize Derived using the most current value for the solar abundance  of iron, $A \log$(Fe)$_{\sun}$ = 7.50 $\pm$ 0.04 \citep{Asplund2009}.}
\end{tablenotes}
\end{threeparttable}
\end{center}
\end{table*}

\subsubsection{Method C -- ARES/MOOG + Schuler linelist}\label{luan}
In the case of the Method C for the unconstrained analysis, the atmospheric parameters (\teff, \logg\ and \mictrb) and metallicity (\feh\footnote{\feh\ = A(\ion{Fe}{1})$_{\star}$ - A(\ion{Fe}{1})$_{\sun}$,
where A(\ion{Fe}{1}) = log [N(\ion{Fe}{1})/N(H)] + 12}) of WASP-13 were derived using the standard spectroscopic method based on the excitation and
ionization equilibrium of \ion{Fe}{1} and \ion{Fe}{2} lines. The \feh\ abundances were normalized on a line-by-line basis to the solar values taken
from \cite{Schuler2011a}. The analysis was done in Local Thermodynamic Equilibrium (LTE) using the 2010 version of MOOG\footnote{Available at
http://www.as.utexas.edu/$\sim$chris/moog.html.} \citep{Sneden1973} and one-dimensional plane-parallel model atmospheres interpolated from the OVER grid of
ATLAS9 models \citep{Castelli2004}.

The line list was adopted from \cite{Schuler2011b} and the equivalent widths (EWs) were measured using the automatic code ARES \citep{Sousa2007}. Effective
temperatures and microturbulence velocities were iterated until the slopes of \feh\ \textit{versus} $\chi$ (the excitation potential of the 
lines) and $\log(EW/\lambda)$ (their reduced equivalent widths) were respectively zero;
i.e., until the individual \feh\ abundances were independent of excitation potential and reduced EWs. Surface gravities were iterated until the \feh\
abundances determined from \ion{Fe}{1} and \ion{Fe}{2} lines were equal. The iteration of the atmospheric parameters was done automatically, using codes
adapted from \cite{Ghezzi2010a}. Any lines with \feh\ abundances that deviated more than 2$\sigma$ from the average were removed and the above iteration was
repeated until convergence was achieved.
The final line list contained 45 \ion{Fe}{1} and 5 \ion{Fe}{2} lines.

The internal uncertainties on the atmospheric parameters were estimated as follows.
The error of the microturbulence was determined by varying this parameter until the slope of \feh\ \textit{versus} $\log(EW/\lambda)$ was equal to its standard
deviation. The uncertainty of the effective temperature was obtained by changing this parameter until the slope of \feh\ \textit{versus} $\chi$ was equal to its
standard deviation. The error of \mictrb\ was also taken into account when calculating the uncertainty of \teff. The error of the surface gravity was obtained
by varying this parameter until the difference between the average [Fe I/H] and [Fe II/H] abundances were equal to the standard deviation of the latter (divided by the square root of 
the number of \ion{Fe}{2} lines). The contribution from \teff\ was also included. Finally, the uncertainty of \feh\ is a combination of the standard deviation
of the [Fe I/H] abundance (divided by the square root of the number of \ion{Fe}{1} lines) and the variations caused by the errors in \teff, \logg, and \mictrb, all added in quadrature.
We note that these are the internal errors of the spectroscopic differential analysis used here and that the real uncertainties (e.g., from the comparison with other similar studies) might be larger. 

In the second case, the surface gravity is fixed to the value determined from the analysis of the 
mean stellar density (\logg\ = 4.10 $\pm$ 0.04),  the other parameters are iterated upon
(with the same line list as above). 
The uncertainties of
\teff\ and \mictrb\ were estimated as described above. The error of the metallicity took into account all four contributions described above, but the influence of the surface gravity
was estimated by varying \logg\ by $\pm$ 1$\sigma$ (i.e., fixing this parameter at the values 4.06 and 4.14) and iterating the atmospheric parameters again.
The larger difference between the new value and the one obtained with \logg\ = 4.10 was taken as the error on \feh\ due to the uncertainty on the surface gravity.
We have observed that this variation of \logg\ has no effect on \feh.

In the third case, we fixed the effective temperature to the value determined from the analysis of the \halpha\ line (\teff\ = 5950 $\pm$ 70 K), and iterated the other parameters
(with the same line list as above). 
The uncertainty of
\mictrb\ was estimated as in the original procedure (with free parameters). The influence of the effective temperature on the errors of the surface gravity and metallicity
was determined by varying \teff\ by $\pm$ 1$\sigma$ (i.e., fixing this parameter at the values 5880 and 6020 K) and iterating the atmospheric parameters again.
The larger differences between the new values and the ones obtained with \teff\ = 5950 K were taken as the errors on \feh\ and \logg\ due to the uncertainty on the
effective temperature and then added in quadrature to the other contributions, giving the final values quoted above.

\subsubsection{Method D -- MOOG/ARES + Sousa linelist}\label{porto}


In the case of Method D, the spectroscopic parameters were derived starting with the automatic measurement of equivalent widths of Fe-I and Fe-II lines with
ARES  \citep{Sousa2007} 
and then imposing excitation and ionization equilibrium using a spectroscopic analysis in LTE with the help
of the code MOOG \citep{Sneden1973} 
and a GRID of Kurucz Atlas 9 plane-parallel model atmospheres \citep{Kurucz1993}. 

The Fe I and Fe II line list is composed of more than 300 lines that were individually tested in high resolution spectra
to check its stability to an automatic measurement with ARES \citep{Sousa2008}.  
The atomic data of the lines were obtained from the Vienna
Atomic Line Database \citep{Kupka1999}  
but the oscillator strength ($\log gf$) of the lines  
were recomputed through an inverse analysis of the Solar spectrum allowing 
this way to perform a differential analysis relatively to the Sun. 
A full description of the method can be found in \citet{Santos2004} and \citet{Sousa2008}. 

We have only reported in Tables~\ref{starchar} and \ref{fixteff} the internal
errors derived from our method. 
Typically, we report a more realisitic
uncertainty that considers the typical dispersion plotted in each comparison of parameters, as presented in 
\citet{Sousa2008}. 
A more complete discussion about the systematic errors generally derived for this spectroscopic method can be 
found in \citet{Sousa2011a}. 
However, given that in this paper we derive the systematic uncertainty from the comparison 
against the resulting spectroscopically-determined parameters from Methods A, B and C, we only report the internal errors for each of the three cases.  

For the constrained cases, the same method was used but fixing each specific parameter (\logg\ and \teff) in the process. We also used the same procedure for the determination of the errors for which the uncertainty in each constrained parameter was considered.





\subsection{Transit Model \label{mcmc}}

The planetary properties were determined using a simultaneous
Markov-Chain Monte Carlo (MCMC) analysis including the WASP
photometry, and the high precision photometry, together with
the radial velocity measurements. A detailed
description of the method is given in \citet{Cameron2007} and
\citet{Pollacco2008}.

Our iterative fitting method uses the following parameters: the epoch
of mid-transit $T_{0}$, the orbital period $P$, the fractional change
of flux proportional to the ratio of stellar to planet surface areas
$\Delta F = R_{\rm pl}^2/R_{\star}^2$, the transit duration $T_{14}$,
the impact parameter $b$, the radial velocity semi-amplitude $K_{\rm
  1}$, the stellar effective temperature \teff\ and metallicity \feh,
the Lagrangian elements \secos\ and \sesin~(where $e$ is the
eccentricity and $\omega$ the longitude of periastron), and the
systematic offset velocity $\gamma$.
The spectroscopically determined \teff\ and \feh\  
presented in the last column of Table~\ref{starchar} with the two reported errors added in quadratures are used 
within our MCMC code as priors in the stellar mass determination from the empirical Torres/Enoch relationship (see below). 
The sum of the $\chi^2$
for all input data curves with respect to the models was used as the
goodness-of-fit statistic.

An initial MCMC solution with eccentricity 
as a free parameter, was explored for WASP-13 deriving a small eccentricity ($e$ = 0.10).
However, the probability that it is a spurious non-circular orbit as defined by
\citet{LucySweeney1971} is 1.0 in agreement with previous circular solutions for the system \citep{Skillen2009,Barros2012}.
Thus, we adopt a circular orbit for the rest of our analysis.  
For the treatment of the
stellar limb-darkening, the four-coefficient law of \citet{Claret2000, Claret2004}
was used with their derived coefficients in the $R$-band for both the JGT and as an approximation for the WASP photometry, which is 
in a $V+R$ passband. 
In the case of the RISE photometry, 
we used the same limb-darkening law, using the coefficients derived specifically
for the RISE passband and CCD response by I. Howarth, following the procedure in \citet{Howarth2011}.

From the parameters mentioned above, we calculate the mass $M$, radius
$R$, density $\rho$, and surface gravity \logg\ of the star (which
we denote with subscript $_{\star}$) and the planet (which we denote
with subscript $_{\rm pl}$), as well as the equilibrium temperature of
the planet assuming it to be a black-body ($T_{\rm
  pl,A=0}$) and that energy is efficiently redistributed from the
planet's day-side to its night-side. 
We also calculate the transit
ingress(egress) times $T_{\rm 12}$($T_{\rm 34})$, and the orbital 
semi-major axis $a$. 
These calculated values and their 1--$\sigma$
uncertainties from our MCMC analysis are presented in
Table~\ref{table:plparams}.  The observed light curves are plotted against the model
light curves with their residuals in Fig.~\ref{lcs}.  

 
  The stellar mass of planet host star has been derived within our MCMC analysis 
 from the empirical 
   \citet{Torres2010} calibration, which is based on the precisely measured masses and radii of eclipsing binary
   stars, and relates \logg, \feh, and \teff\ to the stellar mass and radius.
   However, while \teff\ can be determined with high precision from the 
stellar spectrum (see \S\ref{teff} and \S\ref{starch}), \logg\
   is usually poorly constrained, and thus stellar masses derived from the
   spectroscopic \logg\ can have large uncertainties and can suffer from
   systematics. 
  Thus, our MCMC method derives the stellar mass using the empirical calibration 
as described by \citet{Enoch2010}, which is the based on that by \citet{Torres2010} 
but relies on the directly measured \rhostar, instead of \logg. 
  The stellar density,
   \rhostar, is directly determined from transit light curves and as such
   is independent of the stellar mass, and the effective temperature determined from the
   spectrum \citep{Sozzetti2007,Hebb2009}, as well as of theoretical stellar models
   ($\mpl \ll \mstar$ is assumed; see \citealt{Seager2003}).
The error on the stellar mass that we first derived from the MCMC,
which is based on the empirical relationship, seemed underestimated ($\sim$1\%), 
and thus so are the errors of the other properties that depend on the stellar mass (e.g., orbital separation).
Therefore, we adopt a more realistic uncertainty in the stellar mass from the comparison of stellar evolutionary models
to the observed properties of WASP-13, as described in the paragraph below and is shown in Fig.\~ref{tracks}. 
This uncertainty in the stellar mass  
is included in the final MCMC analysis (see Table~\ref{table:plparams}) and is propagated through all other dependent parameters.  
 
\begin{figure*}[!ht]
\includegraphics*[width=1.0\textwidth]{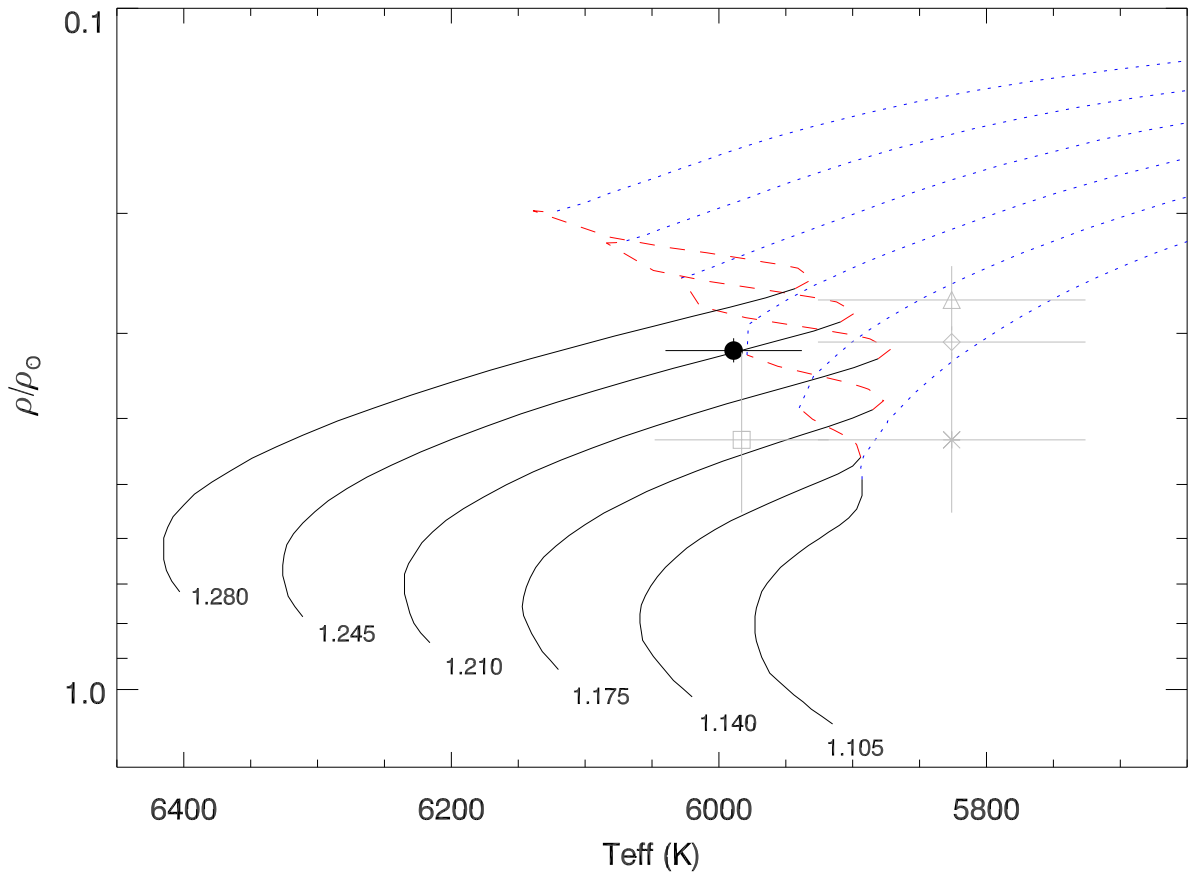}
\caption{\label{tracks} 
Parameters of WASP-13 compared to the Yonsei-Yale stellar evolution models \citep{Demarque2004}.   
The effective temperature and mean stellar density of WASP-13 is plotted as a solid circle.    
The six mass tracks have been interpolated in metallicity to [Fe/H]=+0.06 to match the measured value of WASP-13.  
The mass of each track is labeled at the bottom of each line.  
Different evolutionary phases -- main sequence evolution (solid black), 
overall contraction phase 
(dashed red) and post main sequence (blue dotted) -- are given different line styles.    Previous measurements of WASP-13 are  also plotted in grey -- diamond \citep{Barros2012},  triangle \citep{Southworth2012}, square \citep{Torres2012}, and asterisk  \citep{Skillen2009}.  The best mass for WASP-13 derived from these tracks is 1.245 \msun\ if the star is on the main sequence, but a 1.175 \msun\ star is equally likely if the star is slightly older and has already exhausted hydrogen in its core.     
}
\end{figure*}

Because the planet physical properties depend directly on the stellar ones and to assure the validity
of our MCMC results, we have
derived independently 
the stellar mass from the Yonsei-Yale (Y$^2$) stellar evolutionary models  \citep[see Fig.~\ref{tracks};][]{Demarque2004}. 
Furthermore, this consistency check allows us to estimate the age of the planetary system, 
and to derive a realistic error on the mass of the stellar host.  
We have interpolated the Y$^2$ models considering the 1$-\sigma$ errors in the measured 
\rhostar, and in the spectroscopically determined \feh\ and \teff.  
As shown in Fig.~\ref{tracks}, 
the evolutionary state of WASP-13 is not uniquely determined by the measured stellar properties alone. 
Different mass tracks for the pre-main sequence, main sequence and post-main sequence 
evolutionary phases overlap in the stellar density-effective temperature-metallicity plane at the position of WASP-13.   
We must use additional criteria to identify the most likely mass and age for WASP-13.  First, 
there is no evidence of youth in our WASP-13 data;  the measured lithium abundance, A(Li) = 2.11 $\pm$ 0.08 dex, is consistent with an age of several Gyr 
\citep[see][]{Sestito2005}.  Thus, 
we do not consider the pre-main sequence phase 
as a plausible evolutionary state for WASP-13.
In addition, the measured surface gravity and temperature rule out  
the post-main sequence, red giant phase.  
Therefore, the most likely scenario is that WASP-13 is at the end of its main sequence lifetime and may or may not have reached
the phase of overall contraction before exhausting hydrogen in its core. 
According to these models, the stellar mass is between 1.175, and 1.245 \msun\ depending on the precise phase of evolution, and the age
of the system is between 4-5.5~Gyr.    
Thus, the stellar mass derived from the Y$^2$ theoretical models is consistent with the stellar mass from our MCMC analysis 
(1.187~\msun).  
However, the range of possible stellar masses derived from a single set of evolutionary models is larger than the uncertainty on the stellar mass given by the empirical Enoch relation.  
Therefore, we conservatively adopt a larger uncertainty of $\sigma_{\mstar} = \pm 0.065$ \msun\ 
on this parameter to account for all plausible mass values. 

 \begin{table*}
\begin{center}
\begin{threeparttable}[t]
 \caption{System parameters and $1\sigma$ error limits derived from the MCMC analysis. \label{table:plparams}}
 \begin{tabular}{lccc}
\hline \hline \\ 
Parameter & Symbol & Value & Units \\ \hline \\
 Transit epoch  & $T_0$ & $  5305.62823   \pm  0.00025 $ & days\tnote{$\dagger$}  \\ 
 Orbital period & $P$ & $  4.3530135   \pm  0.0000027$ & days \\
 Planet/star area ratio & $(\rpl/\rstar)^2$ & $  0.00844 \pm  0.00016 $ & \\
 Transit duration & $t_T$ & $  0.1668 \pm  0.0010 $ & days \\
 Impact parameter & $b$ & $  0.603  \pm  0.025 $ & \rstar \\
 Stellar reflex velocity & $K_1$ & $  0.0555 \pm 0.0036  $ & \kms \\
 Center-of-mass velocity & $\gamma$ & $  9.8348   \pm 0.0009$ & \kms \\
 Orbital eccentricity & $e$ & $  0.$ &  fixed  \\
 Orbital inclination & $i$ & $  85.43  \pm  0.29$ & degrees \\
 Stellar density & \rhostar & $  0.306 \pm  0.020 $ & \rhosun \\
 Stellar mass & \mstar & $  1.187  \pm  0.065 $ & \msun \\
 Stellar radius & \rstar & $  1.574   \pm  0.048  $ & \rsun \\
 Orbital semi-major axis & $a$ & $  0.0552 \pm 0.0010  $ & $AU$ \\
 Planet radius & \rpl & $  1.407   \pm  0.052 $ & \rj \\
 Planet mass & \mpl & $  0.500  \pm 0.037 $ & \mj \\
 Planet surface gravity & \logg$_p$ & $  2.764 \pm 0.038 $ & [cgs] \\
 Planet density & \rhopl & $  0.180 \pm  0.020 $ & \rhoj \\
 Planet temperature& \teq & $  1548   \pm   22 $ & K \\
\hline 
\end{tabular}
\begin{tablenotes}
\item[$\dagger$]{\footnotesize Given in BJD$_{\rm TDB}$ -- 2\,450\,000 as defined by \citet{Eastman2010}.}
\end{tablenotes} 
\end{threeparttable}
\end{center}
\end{table*}

\section{Discussion and Conclusions} \label{concl}


\begin{figure*}[ht!]
\includegraphics*[height=0.4\textheight,width=0.93\textwidth]{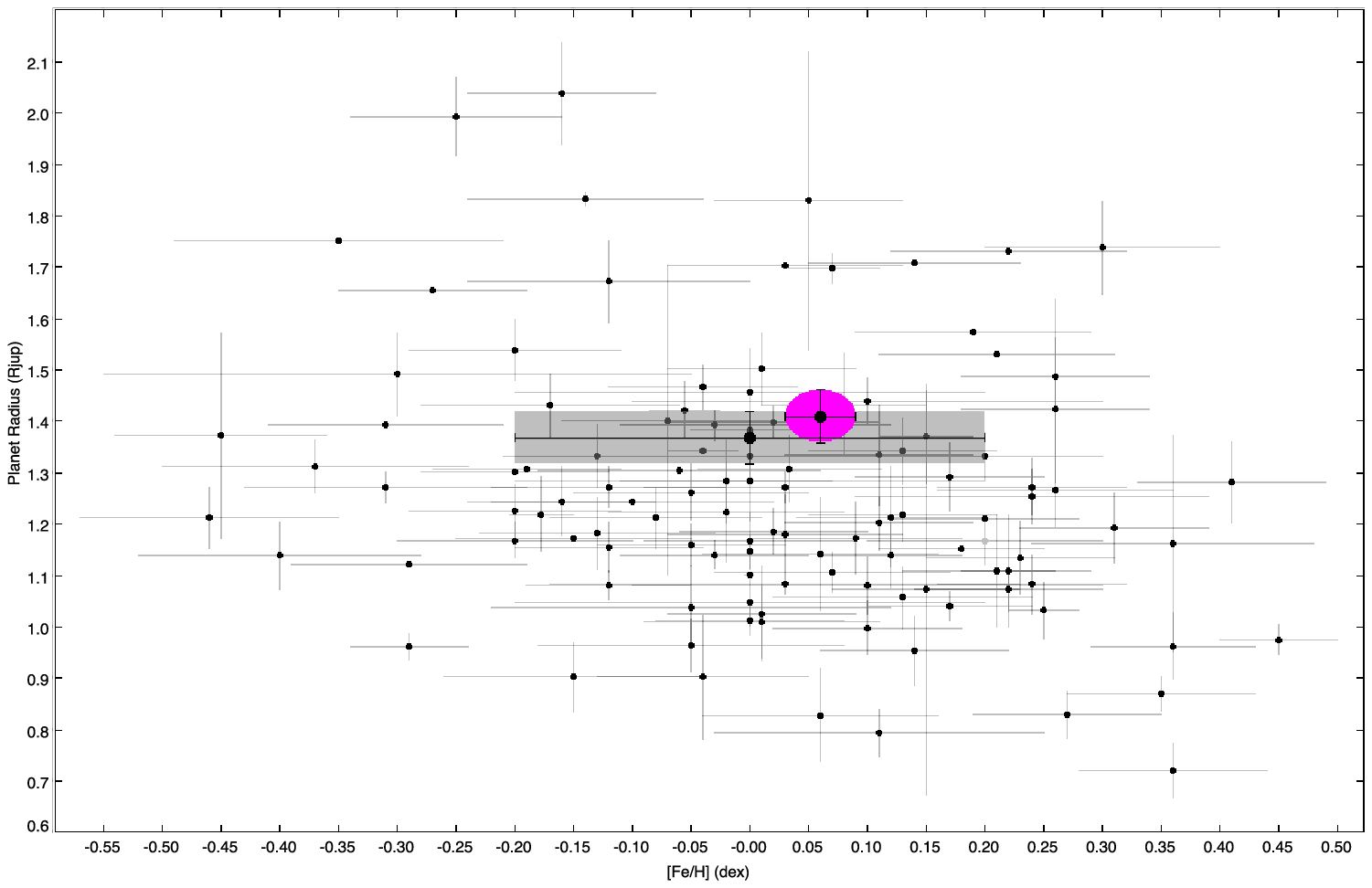}
\includegraphics*[height=0.4\textheight,width=1.0\textwidth]{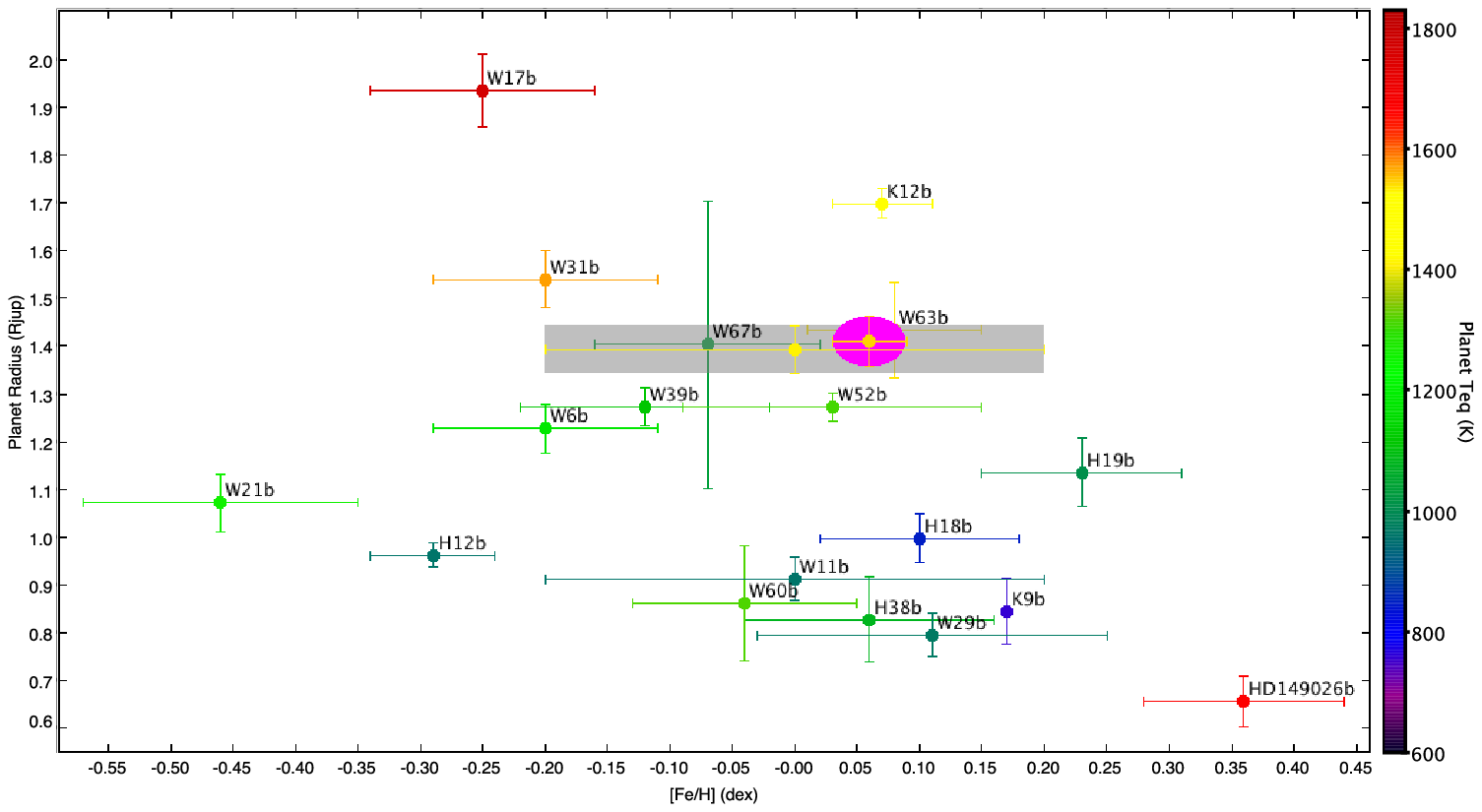}
\caption{\label{figdisc}
{\it Top panel:} 
 \rpl\ versus \feh\ for the bulk of the HoSTS sample taken from {\it exoplanets.org} (2012 Nov 19), and
complemented with the literature.  
A linear regression to the data gives a correlation coefficient of -0.24, which 
supports the previously observed trend \citep[e.g.,][]{Laughlin2011,Enoch2012}.
WASP-13 is marked by the filled circles with the shaded uncertainty areas: 
the grey box is for the \feh\ from \citet{Skillen2009}, and the \rhostar\ from \citet{Barros2012}, 
and the fuchsia oval represents the properties derived in this paper.   
The uncertainties of the other data points are given by the grey crosses.  
{\it Lower panel: }
We show the trend between \rpl, stellar \feh, and the planetary \teq\ (dependent on the stellar irradiation) 
for the HoSTS targets with planets in the Saturn-mass range (0.1 $\le$ \mpl\ $\le$ 0.5 \mj).
Although the two WASP-13 points overlap, the errors on the metallicity are significantly smaller, 
showing the potential of the HoSTS project, by tightening the constraints rendered by the 
known transiting planets, including the identification of any systematics, and assess the validity of the observed trends.  
}
\end{figure*}

In our re-analysis of WASP-13 presented in this paper, we are able 
to more accurately determine the system's physical properties
than in previous studies.
An improvement of 
85\% and 50\% in stellar metallicity and effective temperature, respectively, from the values reported in the discovery paper \citep{Skillen2009} is  
due primarily to the high-quality, high-resolution HIRES spectrum analyzed. 
Moreover, the comparison of the results from the different stellar characterization methods, allow us to 
derive uncertainties based on the internal errors, and including a systematic contribution, based
on the range of derived stellar properties. 
Comparing the stellar spectroscopic properties derived in this paper to those in the literature,
we find a hotter, and slightly more metal-rich star  
(\teff\ = 5989 $\pm$ 51 K; \feh\ = 0.06 $\pm$ 0.03 dex) than  \citet{Skillen2009}, which is 
entirely consistent with the stellar spectroscopic properties of \citet{Torres2012} for WASP-13.  
The stellar mass that we derive (1.187 $\pm$ 0.065~\msun) falls between the recent estimates 
for WASP-13 \citep[1.09 $\pm$ 0.04 and 1.22 $\pm$ 0.12 \msun, respectively;][]{Barros2012,Southworth2012}, 
and above of that initially derived in the discovery paper (1.03 $\pm$ 0.10~\msun).  
Correspondingly, the Saturn-mass planet WASP-13b is found to be also slightly larger and more massive (\rpl\ = 1.407 $\pm$ 0.052 \rj;
\mpl\ = 0.500 $\pm$ 0.037 \mj) with respect to the results from \citet{Skillen2009}, and \citet{Barros2012};
but slightly less massive and smaller than in \citet{Southworth2012}.
Furthermore, we derive a younger planetary system ($\sim$4-5.5 Gyr) than before.  
While WASP-13 has certainly evolved off the zero-age main sequence,
and might have not reached the overall contraction phase,  
it is equally likely that it has undergone contraction and has exhausted hydrogen in its core. 
This uncertainty in the stellar evolutionary stage of WASP-13 
translates into an uncertainty in the stellar mass (see \S\ref{mcmc}, and Fig.~\ref{tracks}),
which has not been typically accounted for.
It is clear from our analysis that both a detailed stellar characterization, and a transit model including 
light curves and radial velocities are necessary to accurately determine the physical properties of planetary systems.   

It must be noted that there is very good agreement among the spectroscopic properties (\teff, \feh, and \logg) of WASP-13 derived from the
four independent methods of stellar characterization for the unconstrained analysis. 
These stellar properties are also consistent with  
the independent measurement of \logg\ constrained from the transit light curve (via \rhostar), and of \teff\ derived from the \halpha\ spectrum. 
Thus, this suggests that there is no significant systematic differences between the methods,
 and that unconstrained spectroscopic analyses give reliable stellar parameters for a star such as WASP-13.
Although seemingly at odds with the conclusions from \citet{Torres2012},
our methodology differs from their spectral synthesis-based analyses (see also \S\ref{sme}).  
Among these differences are the treatment of the microturbulence, the linelist, the sampling of a large parameter 
space in initial parameters, and the convergence criteria.  
Specifically our linelist includes the Na I D region,   
 as well as the region of the Mg b triplet.  
It could be that the biases identified by \citet{Torres2012} due to the 
spectrocopic-\logg\ on the other stellar parameters are not present in our implementation of SME. 
In addition, we have fixed the \teff\ to the value derived from the \halpha\ analysis (\S\ref{teff}) in
the four stellar characterization methods to assess the effect on the other   
spectroscopically-determined stellar properties. In the case of WASP-13, we do not find any
significant differences and the solutions are consistent with the unconstrained analyses.  
Although this is unsurprising given that all the derived temperatures are in agreement, the
effect of fixing \teff\ for different kinds of stars remains to be fully tested. 
With the larger HoSTS sample, a more robust conclusion, as to whether or not to constrain 
the spectroscopic analysis of the planet hosts---with the stellar density from the transit light curve 
or with the \teff\ from a different temperature diagnostic---will be possible  
as a function of analysis method, quality of the data, and/or different stellar and planetary properties.  


Empirical trends between the physical properties
of planet hosting stars and their orbiting exoplanets have been previously identified and  
have been well studied \citep[e.g.,][]{Guillot2006, Fortney2010, Bouchy2010, Baraffe2010, Laughlin2011, Enoch2012, Faedi2011, Demory2011}.
For example, the relationships between planetary radius and stellar metallicity, as well as that between the planetary radius and the 
stellar irradiation,  have been proposed to probe planetary formation and structure.  
According to theoretical models, like those of \citet{Fortney2007}, 
the planetary radius depends on the mass in the core of the planet.   
For example, a core-less Jupiter-mass planet that is dominated by its envelope
has a radius that is several percent larger than a Jupiter-mass planet
with a core-mass of a few tens of \mearth.
At lower planetary masses, like in the Saturn-mass range \citep[0.1 $\le$ \mpl\ $\le$ 0.5;][]{Enoch2012}, 
as the planets become dominated by the core-mass, 
the difference in radii for planets with and without cores seems more pronounced (see lower panel Fig.~\ref{figdisc}) 
The higher stellar metallicity could lead to the formation of planet cores with more metal content, and thus
to smaller planetary radii \citep[e.g.,][]{Guillot2006}.
The planetary radius is also affected by the amount of irradiation received from the host star \citep{Demory2011,Enoch2012,Perna2012}. 
Because of these different contributions, the effect of the stellar metallicity and irradiation on the formation and evolution of 
planets remains unclear.   
The structure of the planet and its environment  
need to be better constrained to be able to account for the diversity of physical properties of the known transiting systems.
Thus, understanding the empirical relationships  that have been previously identified 
between the stellar metallicity and stellar irradiation and the planetary radius 
may expand our knowledge on planetary systems. 

The top panel of Fig.~\ref{figdisc} shows the brightest (V $<$ 14 mag) transiting planets orbiting closest to the host star (P$_{\rm orb}$ $<$ 15 d)
with planetary masses between 0.1 and 12.5 \mj,    
which compose the bulk of the HoSTS sample.
Doing a linear regression including the data uncertainties, we get a correlation coefficient of -0.24 between the 
stellar metallicity and the planetary radius.   
This known anticorrelation is more significant (with a correlation coefficient of -0.37) for the Saturn-mass planets shown in the lower panel
as mentioned above. 
However, there is a strong correlation between the amount of stellar flux received expressed by \teq\ \citep[][see color-scale in lower panel]{Fressin2007} 
which has not been taken into account in the linear regression.  
In this paper, we do not attempt to characterize the observed trends except qualitatively.   
We note the evident 
improvement in the precision of the derived physical properties of the WASP-13 planetary system from our analysis 
(from gray-shaded area to fuchsia-shaded area).  
Our analysis of WASP-13 showcases how the HoSTS project
will allow us to tighten the constraints rendered by the known transiting planets, including the identification of any
systematics, and correlations between the planet and stellar properties, as well as to assess the validity of the known 
empirical trends.

We are continuing to acquire and analyze the high-quality echelle spectra and the 
long-slit spectra for \teff\ determination for the known 
transiting hosts. 
The data products of the HoSTS project are the derivation of a  
homogeneous set of stellar and planetary properties that will
allow us to  
identify any biases in the parameters arising from the analysis methods
and the quality of the data. 
This will enable us to significantly compare the physical properties allowing us
to discover, derive, identify trends among the planetary parameters exploring
in more detail the planetary mass-radius relationship. 


\bibliography{wasp13}
 
\acknowledgments 
{\tiny 
The authors are grateful to the anonymous referee for improving significantly the scientific content of this paper.
Y.G.M.C. and F.F. acknowledge Luca Fossati for fruitful discussions.  
P.A.C. and L.H.H. thank Jeff Valenti, and Eric Stempels for their 
extensive help in running SME and developping the SME implementation presented
in this paper.
Y.G.M.C. acknowledges postdoctoral funding support from the Vanderbilt Office of the Provost, through the Vanderbilt Initiative in Data-intensive Astrophysics (VIDA) and through a grant from the Vanderbilt International Office in support of the Vanderbilt-Warwick Exoplanets Collaboration. L.H.H. and K.G.S. acknowledge National Science Foundation grant AST-1009810. P.A.C. and K.G.S. acknowledge National Science Foundation grant AST-1109612. 
This work was supported by the European Research Council/European Community under the FP7 through Starting Grant agreement number 239953. 
S.G.S. is supported by the grant SFRH/BPD/47611/2008 from FCT (Portugal).
L.G. acknowledges financial support provided by the PAPDRJ CAPES/FAPERJ Fellowship. 
N.C.S. also acknowledges the support from Funda\c{c}\~{a}o para a Ci\^encia e a Tecnologia (FCT) through program 
Ci\^encia 2007 funded by FCT/MCTES (Portugal) and POPH/FSE (EC), and in the form of grant reference PTDC/CTE-AST/098528/2008. 
The INT is operated on the island of La Palma by the Isaac Newton Group in the Spanish Observatorio del Roque de los Muchachos of the Instituto de Astrofísica de Canarias.
 The data presented herein were obtained at the W.M. Keck Observatory, which is operated as a scientific partnership among the California Institute of Technology, the University of California and the National Aeronautics and Space Administration. The Observatory was made possible by the generous financial support of the W.M. Keck Foundation. The authors wish to recognize and acknowledge the very significant cultural role and reverence that the summit of Mauna Kea has always had within the indigenous Hawaiian community.  We are most fortunate to have the opportunity to conduct observations from this mountain.
} 
\end{document}